\documentclass[10pt,fleqn,a4paper]{article}
\pdfoutput=1
\usepackage{graphicx}
\usepackage{epsfig,cite}
\usepackage{amssymb}
\usepackage{amsmath}
\usepackage{color}
\usepackage{amstext,alltt,setspace}
\usepackage{amsbsy}
\usepackage{array}
\usepackage{booktabs, makecell}
\usepackage{hyperref}
\usepackage{slashed}
\usepackage{url}
\usepackage{geometry}
\begin{document}

\begin{flushright}
  TIF-UNIMI-2020-2\\
CAVENDISH-HEP-20-01\\
\end{flushright}

\vspace*{.2cm}

\begin{center}
  {\Large \bf{Why $\alpha_s$ Cannot be Determined from Hadronic
      Processes without Simultaneously Determining the Parton Distributions}}
\end{center}

\vspace*{.7cm}

\begin{center}
  Stefano Forte$^{1}$ and Zahari Kassabov$^2$
  \vspace*{.2cm}

{  \noindent
      {\it
        ~$^1$ Tif Lab, Dipartimento di Fisica, Universit\`a di Milano and\\ 
        INFN, Sezione di Milano,
        Via Celoria 16, I-20133 Milano, Italy\\
        ~$^2$Cavendish Laboratory, University of Cambridge, Cambridge CB3
0HE, United Kingdom\\}}

      \vspace*{3cm}

      {\bf Abstract}
\end{center}

{\noindent We show that any determination of the strong coupling
  $\alpha_s$ from a process which depends on parton
  distributions, such as hadronic processes or deep-inelastic
  scattering, generally does not lead to a correct result unless
  the parton distributions
  (PDFs) are  determined simultaneously along with $\alpha_s$. We
  establish the result by first showing an explicit example, and then
  arguing that the example is representative of a generic situation
  which we explain using models for the shape of equal $\chi^2$
  contours in the joint space of $\alpha_s$ and the PDF parameters.}

\pagebreak

\section{The determination of $\alpha_s$ in hadronic processes}
\label{sec:intro}

The value of the strong coupling $\alpha_s$ has been routinely
determined from a variety of processes which involve hadrons in the
initial state, both in electroproduction and hadroproduction.
The current PDG average~\cite{Tanabashi:2018oca} includes two
different classes of such 
determinations. One is from ``DIS and PDF fits'':
in these determinations the value of $\alpha_s$ is determined
together with a set of parton distributions (PDFs) from a more or less wide
set of data and processes, ranging from deep-inelastic scattering
(DIS) to  hadron collider processes  (such as
Drell-Yan, top, and jet
production).

The other is from single hadronic processes: specifically top  pair
production~\cite{Chatrchyan:2013haa,Klijnsma:2017eqp, Sirunyan:2018goh}, and jet
electroproduction~\cite{Andreev:2017vxu}. Several more determinations of
$\alpha_s$ from one process have been presented, such as for
instance jet production~\cite{Britzger:2017maj, Andreev:2014wwa,
Khachatryan:2016mlc, Affolder:2001hn, Chekanov:2006yc, Abazov:2009nc,
Malaescu:2012ts, Khachatryan:2014waa}, multijets~\cite{Andreev:2014wwa, ATLAS:2015yaa,
Aaboud:2017fml, Aaboud:2018hie, Chekanov:2005ve, Chatrchyan:2013txa,
Abazov:2012lua, CMS:2014mna, Andreev:2016tgi} and $W$ and $Z$
production~\cite{dEnterria:2019aat}. In these determinations, 
PDFs are taken from a pre-existing set, rather than being determined
along with $\alpha_s$. The value of $\alpha_s$ is then found by
determining the likelihood of the new data as a function of $\alpha_s$
--- crudely speaking, by
computing the $\chi^2$ to  the new data 
of the theoretical prediction which  corresponds to a
variety of values of $\alpha_s$, and determining the minimum of the
parabola (though in practice when various parametric uncertainties
have to be properly kept into account the procedure is rather more
elaborate, see e.g. Ref.~\cite{Klijnsma:2017eqp}). The theoretical prediction is in turn obtained for each
value of $\alpha_s$ by combining the matrix element computed with the
given $\alpha_s$ value with the PDF set that corresponds to that
$\alpha_s$ value. This is of course necessary because PDFs strongly
depend on $\alpha_s$, so  a consistent calculation requires the
use of PDFs corresponding to that value. All major PDF sets are
available for a variety of $\alpha_s$ values, and thus this poses no
difficulty in practice.

Here we will show that this apparently straightforward and standard
procedure may lead
to an incorrect determination of $\alpha_s$, and we will argue that
this is in fact a generic situation. The difference between this and
the true best fit $\alpha_s$ can be very
substantial, and specifically much larger than the statistical
accuracy of the $\alpha_s$ determination: as we shall see, this in
fact reflects a conceptual flaw in the procedure.

The reason for this can be understood by viewing the $\chi^2$ as
a simultaneous function of $\alpha_s$ and the PDF parameters. Any
given existing 
PDF set then traces a line in such space (the ``best-fit line'',
henceforth):
for each value of $\alpha_s$
there is a set of best-fit PDF parameters, which corresponds to a
point in PDF space. The standard procedure seeks for the minimum of
the $\chi^2$ in this subspace. This disregards the fact that the
true minimum generally corresponds to a different point in (PDF,
$\alpha_s$) space, which also accommodates the new data~\cite{Kassabov:2018bav}.

One could naively argue that the standard procedure is
correct, because what one is really doing is determining the best
$\alpha_s$ value for the new process subject to the constraint that
PDFs describe well the (typically very large) set of data used to 
determine them. And surely -- the naive argument goes --- the minimum of
$\alpha_s$ anywhere other than on the best-fit line must
correspond to a worse description of the world data?
It actually turns out that this is incorrect:
there exist points in (PDF, $\alpha_s$) space  for which the 
value of the $\chi^2$ for the new process is lower than any value
along the best-fit line, yet, somewhat counter-intuitively,
the value of the $\chi^2$ for the world data is also
  lower.

Moreover, the value of $\alpha_s$ corresponding to these configurations may,
and in general will, differ substantially from the one obtained using the
standard procedure, and in particular it will be closer to the value obtained by
simultaneously fitting $\alpha_s$ and PDFs to a global
dataset. Therefore, the standard procedure  leads to
a distorted answer, and it inflates artificially the dispersion of $\alpha_s$
values obtained from different processes.

We establish this result by first providing an explicit example
in which this happens.
Namely, we consider the dataset used for
the NNPDF3.1~\cite{Ball:2017nwa} PDF determination. We then study
the $\chi^2$ for the subset of data corresponding to the
$Z$  transverse-momentum ($p_t$)  distribution, and
determine the best-fit value of $\alpha_s$ from this $Z$ $p_T$
distribution along the best-fit line corresponding to the global fit dataset.
We
then exhibit a specific set of PDFs corresponding to
a rather different value
of $\alpha_s$, and such that
the $\chi^2$ is better both for  the $Z$ $p_T$ distribution,
and for the rest of the dataset. This means that there exists at least
one point in  (PDF, $\alpha_s$) space such the value of the $\chi^2$
for the  $Z$ $p_T$ is
better than any value along the best-fit line, and that there is no
reason not to consider this as a better fit than the result at the
best-fit $\alpha_s$ along the best-fit line, because the agreement
with the world data is also better than that at the minimum on the
best-fit line. 

We will understand the reason for this result by providing models
for the shape of the $\chi^2$ contours both for the world data and the
new experiment in the joint  (PDF, $\alpha_s$) space. Specifically, we
explain  that this situation may arise both in the case in
which the new data may provide an independent determination of
$\alpha_s$ and the PDFs of its own, and in the
case in which the new data do not determine $\alpha_s$ and the PDFs
independently. This then covers the typical realistic scenarios in
which the new data  only constrain (or determine) a subset of PDFs:
e.g. in the case of the $Z$ $p_T$ distribution considered above, the
gluon. In this latter, common case we will see that the
value of $\alpha_s$ obtained through the standard procedure leads to
an artificially large dispersion of $\alpha_s$ values: better-fit
points in  (PDF, $\alpha_s$) generally lead to $\alpha_s$ values which
are closer to the global best fit.

\section{An explicit example: the $Z$ transverse momentum
  distribution} 
\label{sec:zpt}

We  provide an explicit example of the situation we described in the
introduction.  We  consider the $\chi^2$ values for both a global ``world''
dataset, and the dataset for a particular process $P$, as a function of
$\alpha_s$.  Given a fixed value of $\alpha_s$, the value of
$\chi^2$ also depends on the PDF
set which is being used. As $\alpha_s$ is varied, there is a PDF set which
corresponds to the global best fit: this PDF set defines a line in (PDF,
$\alpha_s$) space which we call the best-fit line. We  call
$\chi^2_g(\alpha_s)$ the value of the $\chi^2$ for the global dataset, as a
function of $\alpha_s$, along this best-fit line.

We now consider the $\chi^2$ for process $P$:
We denote by   ${\chi^r_{P}}^2(\alpha_s)$ the value
of the $\chi^2$ for process $P$ as a function of $\alpha_s$, along
this same best-fit line in (PDF, $\alpha_s$) space. We call this the
restricted $\chi^2$ for process $P$.  This means that  this restricted
${\chi^r_{P}}^2(\alpha_s)$ is
found using the value
$\alpha_s$ of the strong coupling, but the PDF set which corresponds
to the {\it global} best fit. So $\chi^2_g(\alpha_s)$ and
${\chi^r_P}^2(\alpha_s)$  are determined using the same $\alpha_s$
and the same PDF set: that which corresponds to the global
best fit.
Note that, for any value of
$\alpha_s$,  this restricted
${\chi^r_P}^2(\alpha_s)$ is not in general  the
lowest $\chi^2$ value for process $P$ that can be found with the given
value 
$\alpha_s$ of the strong coupling --- the PDFs are optimized for the
global dataset, not for process $P$.
This is unlike the global
$\chi^2_{g}(\alpha_s)$, in which (by definition) for each
$\alpha_s$ choice, the PDF set
is always chosen as the corresponding global best-fit PDF set.

Now, the standard procedure determines $\alpha_s$ from process $P$ as
the minimum of ${\chi^r_P}^2(\alpha_s)$: namely, as the value of
$\alpha_s$ which minimizes $\chi^r_P$, the restricted $\chi^2$ for
process $P$, evaluated along the best-fit line. We call this value of
$\alpha_s$, determined using the standard procedure,
${\alpha_0^r}^{P}$: the restricted best-fit value of $\alpha_s$, and the
corresponding PDF set the restricted best-fit PDF set for process $P$.

We now show that this restricted  ${\alpha_0^r}^{P}$ cannot be viewed as
the value of $\alpha_s$ determined by process $P$. We do this 
by exhibiting a
 point in (PDF, $\alpha_s$) space which does not lie along the
best-fit line,  i.e. such that the PDFs do not correspond to the global best
fit, such that
 $\alpha_s\not={\alpha_0^r}^P$, and such that both
the $\chi^2$ for the individual dataset, and for the global dataset, are
respectively better than the restricted ${\chi^r_P}^2({\alpha_0^r}^P)$  and
$\chi^2_{\rm   g}({\alpha_0^r}^P)$.
This is thus a better fit to both process $P$ and the global
dataset than the restricted best fit, so 
there is no sense in which the restricted best-fit
${\alpha_0^r}^P$ --- which would be
the ``standard'' answer --- can be considered   the $\alpha_s$ value
determined by process $P$.

Our construction is based on a previously published determination of
$\alpha_s$ by the NNPDF collaboration~\cite{Ball:2018iqk}, in which
  the strong coupling is determined together with a set of parton
  distributions based on a global dataset which is very close to that
  used for the NNPDF3.1~\cite{Ball:2017nwa}.
  This $\alpha_s$
  determination, which we now briefly summarize for completeness,
  builds upon the previous NNPDF methodology for PDF
  determination, in which PDFs are determined as a Monte Carlo set of
  PDF replicas, each of which is fitted to a replica of the underlying
  data. Note that, in this $\alpha_s$
  determination,  the PDFs and $\alpha_s$ are fitted
  simultaneously. This is unlike the case of previous
  determinations~\cite{Ball:2011us} in which  PDFs were
  determined for a variety of $\alpha_s$ values, and then the best fit
  was sought by looking at the likelihood profile of the best fit as a
  function of $\alpha_s$. Whereas the two methodologies lead (if
  correctly implemented) to the same best-fit $\alpha_s$ value,
  simultaneous minimization ensures a more accurate determination of
  the uncertainty involved, as explained in Ref.~\cite{Ball:2018iqk},
  essentially because it determines the likelihood contours in
  (PDF, $\alpha_s$) space, rather than just the likelihood line
  corresponding to the best-fit PDF for each $\alpha_s$ value.

  The way this is accomplished in Ref.~\cite{Ball:2018iqk} within the
  NNPDF methodology is by fitting each data replica
  several times for a number of different values of $\alpha_s$, thereby
  providing a correlated ensemble of PDF replicas, in which to each
  data replica corresponds a PDF replica for each value of
  $\alpha_s$. Namely, for the $k$-th data replica $D^{(k)}$, a PDF
  replica is found by   determining
  the
  set of PDF parameters $\theta^{(k)}$ which minimize the $\chi^2$:
\begin{align}
\theta^{(k)}(\alpha_s)&=\mathrm{argmin}_\theta\left[\chi^2(\theta,D^{(k)},\alpha_s)\right]\,
,
\label{eq:thetakdef}
\end{align}
where by $\mathrm{argmin}_\theta$ we mean that the minimization is
performed with respect to  $\theta$ for fixed  $D^{(k)}$ and $\alpha_s$.
It is then possible to compute the $\chi^2$ for the $k$-th data
replica as $\alpha_s$ is varied:
\begin{equation}
\chi^{2(k)}(\alpha_s)=\chi^{2} \left(\alpha_s, \theta^{(k)}(\alpha_s),D^{(k)} \right)\, .
\label{eq:chi2kdef}\end{equation}
We thus find an ensemble of parabolas $\chi^{2(k)}(\alpha_s)$,
one for each data replica.
The best-fit $\alpha_s$ for the $k$-th data replica corresponds to the
minimum along the $k$-th parabola:
\begin{equation}
\alpha_s^{(k)}=\mathrm{argmin}\left[\chi^{2(k)}(\alpha_s)\right].
\label{eq:alphakdef}\end{equation}
In the NNPDF approach, the best-fit PDF value is
the average of the PDF replica sample; similarly the best-fit
$\alpha_s$ is  determined  averaging the $\alpha_s^{(k)}$
values. We refer to Ref.~\cite{Ball:2018iqk} for further details,
specifically on the dataset. Here we will use the NNLO PDF replicas
determined in that reference as our baseline.

We can now consider any particular process $P$ entering these global PDF
determination, and ask ourselves what is the $\alpha_s$ value
corresponding to process $P$. The ``standard'' answer would be to simply
consider the ensemble of best-fit PDFs determined in the global fit,
and compute again $\chi^{2(k)}(\alpha_s)$ but now only including
process $P$ in the computation of the $\chi^2$. We then get another
set of parabolas
\begin{equation}
{\chi^r}^{2(k)}_P(\alpha_s)=\chi^{2} \left(\alpha_s, \theta^{(k)}(\alpha_s),D_P^{(k)} \right)
\ , 
\label{eq:chi2kdefP}\end{equation}
where only the data $D_P$ for process $P$ have been used. Note that
these are restricted $\chi^2$ parabolas, because the PDF parameters
$\theta^{(k)}(\alpha_s),$ are those found in Eq.~(\ref{eq:thetakdef}),
by minimizing the global $\chi^2$.
The minima
\begin{equation}
{{\alpha^r_s}^{(k)}}_{P}=\mathrm{argmin}\left[{\chi_P^r}^{2(k)}(\alpha_s)\right]
\label{eq:alphakmin}\end{equation}
now give an ensemble of restricted best-fit $\alpha_s$ values for
process $P$. Their average is then the restricted best fit for this
process.

\begin{figure}[t]
\begin{center}
  \includegraphics[width=0.49\textwidth]{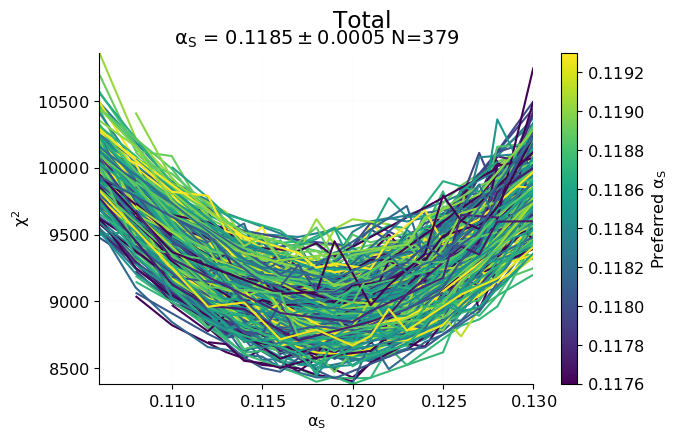}
  \includegraphics[width=0.49\textwidth]{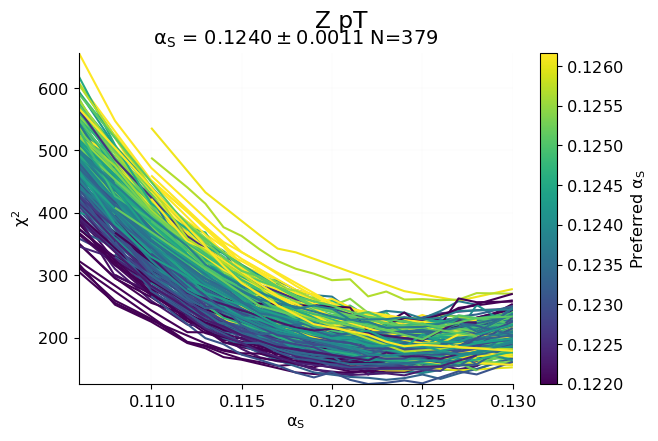}
  \caption{\small The $\chi^2$ profiles for each of the
 data replicas  used for
    the NNLO determination of $\alpha_s(m_Z)$ of
    Ref.~\cite{Ball:2018iqk}. Both the profiles for the total dataset (left),
    and for the $Z$ $p_T$ distribution (right) are shown.
   }
    \label{fig:nnlo_parabolas}
\end{center}
\end{figure}

\begin{figure}[t]
\begin{center}
  \includegraphics[width=0.49\textwidth]{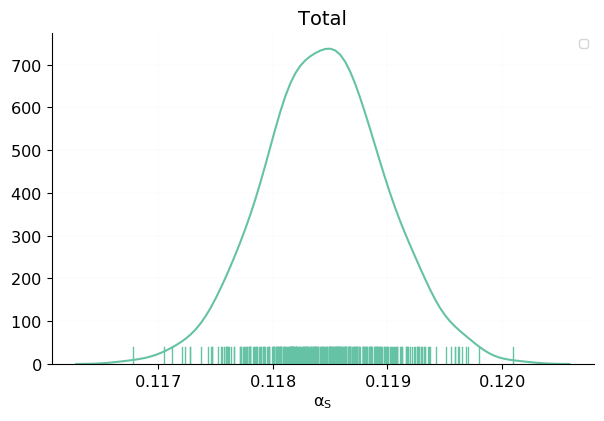}%
  \includegraphics[width=0.49\textwidth]{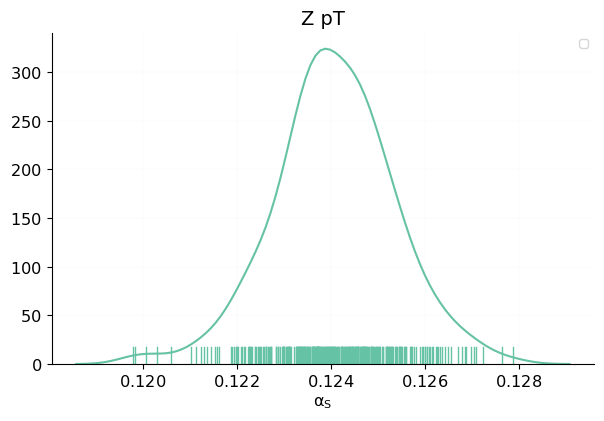}
  \caption{\small
The probability distributions for the best-fit values $\alpha_s^{(k)}$
Eq.~(\ref{eq:alphakdef}) and ${{\alpha_s^r}^{(k)}}_{P}$  
Eq.~(\ref{eq:alphakmin}) respectively
for the global dataset (left) and the $Z$
$p_T$ distribution (right).
Each marker indicates the   value corresponding to
 each individual parabola of Fig.~\ref{fig:nnlo_parabolas}.
    \label{fig:as_distributions}}
\end{center}
\end{figure}

In Fig.~\ref{fig:nnlo_parabolas} we show the parabolas corresponding
both to the global fit (left) and to the $Z$ $p_T$ distribution (right). The
corresponding ensemble of values of $\alpha_s$ is shown in
Fig.~\ref{fig:as_distributions}. From these we find that the global
best-fit value of $\alpha_s(M_Z)$ is 
\begin{equation}\label{eq:globalpha}
  \alpha_s(M_z)=\alpha_0^g=0.1185\pm0.0005,
  \end{equation}
while the restricted best fit is, for the $Z$ $p_T$ distribution,
\begin{equation}\label{eq:zptalpha}
\alpha_s(M_Z)=  {\alpha_0^r}^{Z\, p_t}=0.1240\pm0.0015.
\end{equation}
In both cases, the central value and
uncertainty are respectively the mean and standard deviation computed
over the replica sample, in the first cases for the global best fit
Eq.~(\ref{eq:alphakdef}) and
in the latter case for the restricted  best fit Eq.~(\ref{eq:alphakmin}) for
each replica.

\begin{table}[h]
\begin{center}
  \scriptsize
      \renewcommand{\arraystretch}{1.10}
\begin{tabular}{lcccc}
  \toprule
 Dataset & \thead{$N_{\rm dat}$} & \thead{default \\ $\alpha_s=0.120$ }&
 \thead{reweighted \\ $\alpha_s=0.120 $ } & \thead{default \\ $\alpha_s=0.124$}    \\
\midrule
NMC    & 325 &1.315 &1.337 &1.341 \\    
SLAC    & 67  &0.6787 &0.6994 &0.7198\\ 
BCDMS   & 581  &1.232 &1.282 &1.270\\ 
CHORUS  & 832 &1.176 &1.200 &1.249\\ 
NuTeV dimuon  & 76  &0.9229 &0.9125 &0.9900\\ 
\midrule
HERA I+II inclusive  & 1145  &1.263 &1.271 &1.288\\ 
HERA $\sigma_c^{\rm NC}$  & 37  &1.533 &1.538 &1.748\\ 
HERA $F_2^b$   & 29 &1.299 &1.282 &1.247\\ 
\midrule
DY E866 $\sigma^d_{\rm DY}/\sigma^p_{\rm DY}$  & 15  &1.019 &1.020 &1.048\\ 
DY E886 $\sigma^p$  & 89 &0.4322 &0.4221 &0.4477\\ 
DY E605  $\sigma^p$  &85  &1.001 &1.080 &1.020\\ 
\midrule
CDF $Z$ rap  & 29  &1.442 &1.558 &1.419\\  
\midrule
D0 $Z$ rap  & 28  & 0.5990&0.6381 &0.5996\\ 
D0 $W\to e\nu$  asy   & 8 &2.794 &2.860 &2.979\\ 
D0 $W\to \mu\nu$  asy  & 9  &1.594 &1.610 &1.629\\ 
\midrule
ATLAS $W,Z$  & 30  & 0.8957&0.8912 &0.9623\\ 
ATLAS high-mass DY 7 TeV  & 5 &1.819 &1.845 &1.904\\ 
ATLAS low-mass DY 2011   & 6 &1.123 &1.060 &1.605\\ 
ATLAS $W,Z$ 7 TeV 2011   & 34 &2.149 &1.889 &2.289\\ 
ATLAS jets 2010 7 TeV   & 31 &1.478 &1.513 &1.479\\ 
ATLAS $\sigma_{tt}^{\rm tot}$   & 3 & 0.8520& 0.7088& 3.503\\ 
ATLAS $t\bar{t}$ rap & 10  &1.555 &1.339 &2.214\\ 
\midrule
CMS $W$ asy 840 pb   & 11 &0.7858 &0.7804 &0.8083\\ 
CMS $W$ asy 4.7 fb   & 11  &1.762 &1.749 &1.763\\ 
CMS Drell-Yan 2D 2011   & 110 &1.264 &1.332 &1.294\\ 
CMS $W$ rap 8 TeV  &22  &1.010 &1.068 &1.177\\ 
CMS jets 7 TeV 2011   & 133  &0.9766 &1.026 &1.034\\ 
CMS $\sigma_{tt}^{\rm tot}$  & 3 &0.9832 &0.5803 &5.489\\ 
CMS $t\bar{t}$ rap  &10  &1.035 &1.036 &1.069\\ 
\midrule
LHCb $Z$ 940 pb   & 9  &1.595 &1.773 & 1.568\\ 
LHCb $Z\to ee$ 2 fb   & 17  &1.156 &1.184 &1.274\\ 
LHCb $W,Z \to \mu$ 7 TeV   & 29  &1.793 &2.034 &1.894\\ 
LHCb $W,Z \to \mu$ 8 TeV  & 30  &1.440 &1.617 &1.722\\ 
\midrule
{\bf Global dataset}   & 3979 &{\bf 1.212} & {\bf 1.235}&{\bf 1.262}\\ 
\bottomrule
\end{tabular}
\caption{\small
\label{tab:chi2taball}
The values of $\chi^2/N_{\rm dat}$ for the experiments included in the
best global fit with $\alpha_s=0.120$, compared to results obtained
when $\alpha_s=0.124$, or when the $Z$ $p_T$ data are given a large weight
and $\alpha_s=0.120$. The number of datapoints is also given in each
case. The full description of the datasets, including
data selection, cuts, and references is  given
in Ref.~\cite{Ball:2017nwa} where the same data coding is used.
}
\end{center}
\end{table}

\begin{figure}[t]
\begin{center}
  \includegraphics[width=0.49\textwidth]{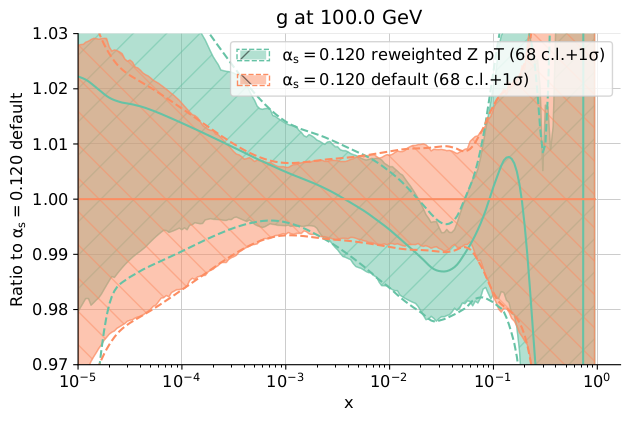}%
  \includegraphics[width=0.49\textwidth]{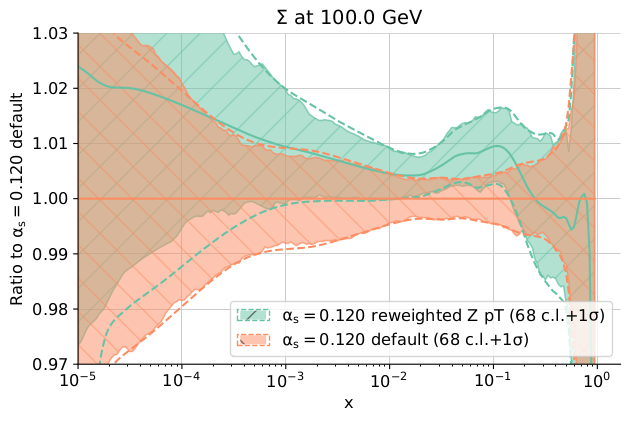}
  \caption{\small
Comparison between the gluon (left) and quark singlet (right) PDFs in
the default global PDF determination (orange, lower band at low $x$)
and in a PDF determination in
which the $Z$ $p_T$ data receive a large weight (green, higher band at
low $x$), shown as a ratio to
the former.
    \label{fig:pdfcomp}}
\end{center}
\end{figure}

We now show that the naive conclusion that the value
Eq.~(\ref{eq:zptalpha}) of $\alpha_s$ is the value of the strong
coupling determined by the $Z$ $p_T$ distribution rests on shaky ground. To show
it, we perform a new PDF determination in
which the $Z$ $p_T$ are now given a large weight in the $\chi^2$, and
which is otherwise identical to the default determination. This PDF
determination is performed for a single value of $\alpha_s(M_z)=0.120$, a
value intermediate between the restricted best-fit  ${\alpha_0^r}^{Z\, p_t}$
Eq.~(\ref{eq:zptalpha}) and the global best-fit
  $\alpha_0^g$ Eq.~(\ref{eq:globalpha}). Specifically
the contribution of the $Z$ $p_T$ data to the total $\chi^2$ has been
multiplied by a factor $w=32$. This factor is chosen so that the
contribution of the  $Z$ $p_T$ data is roughly equal to that of all
the other data.
The gluon and total quark singlet PDFs obtained in this way are
compared in Fig.~\ref{fig:pdfcomp} to  the default PDFs for the same value of
$\alpha_s(M_Z)=0.120$; $\chi^2$ values for the global dataset are collected in
Table~\ref{tab:chi2taball}, while in Table~\ref{tab:chi2tabzpt} $\chi^2$  values
for the $Z$ $p_T$ data and the global dataset are compared. The gluon
is shown because it is the PDF which is most affected by the $Z$ $p_T$
data, and the singlet is also shown because it mixes with the gluon
upon perturbative evolution.

\begin{table}[h]
\begin{center}
  \scriptsize
      \renewcommand{\arraystretch}{1.10}
\begin{tabular}{lccccc}
\toprule 
\thead{Dataset} & \thead{$N_{\rm dat}$}& \thead{default \\ $\alpha_s=0.120$} &
\thead{default \\ $\alpha_s=0.120$, no $Z$ $p_T$} & \thead{reweighted \\
$\alpha_s=0.120 $} & \thead{default \\ $\alpha_s=0.124$ }    \\
\midrule
ATLAS $Z$ $p_T$ 8 TeV $(p_T^{ll},M_{ll})$      &44  &0.9776 & 0.9775 &0.9380 &0.9559 \\    
 ATLAS $Z$ $p_T$ 8 TeV $(p_T^{ll},y_{ll})$       &48  &0.9999 &1.071  & 0.7455&0.8568 \\    
  CMS $Z$ $p_T$ 8 TeV $(p_T^{ll},M_{ll})$    &28  &1.308 &1.299  &1.403 &1.357 \\    
\midrule
All $Z$ $p_T$      &120  &1.056 & 1.085 &0.9635 &1.011 \\    
\midrule
{\bf Global dataset}   & 3979 &{\bf 1.212} & {\bf 1.211 } & {\bf 1.235}&{\bf 1.262}\\ 
\bottomrule
\end{tabular}
\caption{\small
\label{tab:chi2tabzpt}
Same as Table~\ref{tab:chi2taball}, but now comparing the
values of $\chi^2/N_{\rm dat}$ for the $Z$ $p_T$ distributions
and the global dataset. The values for the global dataset are the same
as in  Table~\ref{tab:chi2taball}, while the values for the total $z$
$p_T$ are obtained by combining the three datasets listed in this
table. We also include values for a global fit from which the $Z$
$p_T$ data have been excluded.
}
\end{center}
\end{table}

The logic behind this procedure is that by giving more weight to this
data we obtain a set of PDFs which provide a better fit to them: so
we expect the value of $\chi^2$ for the $Z$ $p_T$ data to be better than
that which would be obtained by taking the default best-fit PDF
set for the same $\alpha_s$ value.
In fact it turns out that the value of the $\chi^2$ thus obtained for
the $Z$ $p_T$ data is also better than the value
${\chi^r_P}^P(0.124)$ which corresponds to the best fit along the
global best-fit line (see Table~\ref{tab:chi2tabzpt}).
This means that the value $\alpha_s(M_Z)=0.120$ is a better fit to the  $Z$
$p_T$ than the value Eq.~(\ref{eq:zptalpha}) corresponding to the best
fit along the best-fit line.

As discussed in the introduction one
might object to the conclusion  that $\alpha_s(M_Z)=0.120$ might be
a better $\alpha_s$ from  $Z$ $p_T$: on the grounds that
the PDF which we obtained
thus are not compatible with the rest of the global dataset given that
they do not correspond to the global best fit. 
However (see again Table~\ref{tab:chi2tabzpt})
the value of $\chi^2$ for the {\it global}
dataset obtained using these PDFs is also better than the value of
${\chi^2}_{\rm g}(0.124)$: hence with $\alpha_s(M_Z)=0.120$ and these PDFs
one gets a
better fit to the $Z$ $p_T$ data than with $\alpha_s(M_Z)=0.124$, while
also better fitting the world data. As it is clear from
Fig.~\ref{fig:pdfcomp}, the PDFs that best reproduce the $Z$ $p_T$ data,
though compatible within uncertainties with the global fit, differ
from them by an amount which is sufficient to considerably improve the
description of the $Z$ $p_T$ data. Indeed, they  lead to an improvement of
their $\chi^2$ value by almost 10\% in comparison to that of the
global fit with the same $\alpha_s(M_Z)=0.120$ value, at the cost of only a
small deterioration of the $\chi^2$ of the global fit, by about 2\%.

\begin{figure}[t]
\begin{center}
  \includegraphics[width=0.49\textwidth]{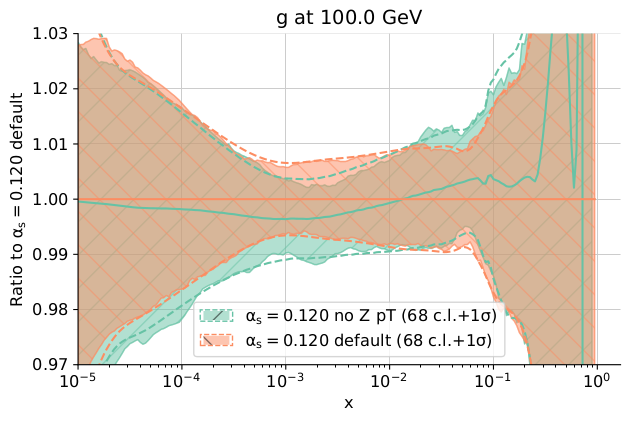}%
  \includegraphics[width=0.49\textwidth]{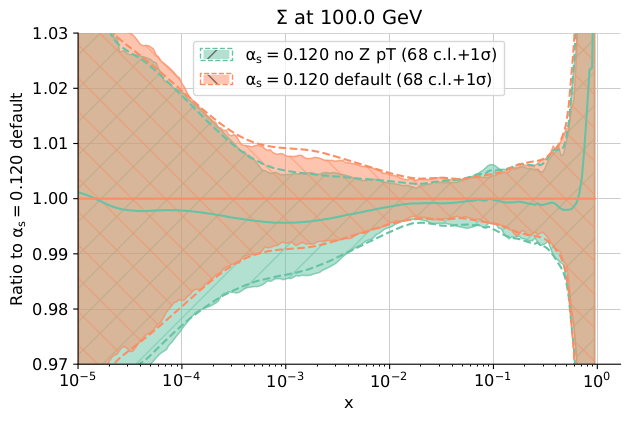}
  \caption{\small
    Same as Fig.~\ref{fig:pdfcomp}, but now comparing the global fit
    (same as shown in Fig.~\ref{fig:pdfcomp}) to a global fit from
    which the $Z$ $p_T$ data have been removed, shown as a ratio to
    the former.
    \label{fig:pdfcompnozpt}}
\end{center}
\end{figure}
The conclusion that 
the restricted best-fit value $ {\alpha_0^r}^{{Z\, p_t}}$
Eq.~(\ref{eq:zptalpha})  is the value of the strong
coupling determined by the $Z$ $p_T$ distribution is thus difficult to
defend:
with $\alpha_s(M_Z)=0.120$ we can fit better both the  $Z$ $p_T$ and the
global dataset, provided the PDFs are suitably readjusted.
It is perhaps worth stressing that the effect that we
are demonstrating is large in comparison to uncertainties. Indeed, the
 global best fit Eq.~(\ref{eq:globalpha}) differs by almost four
 standard deviations from the restricted best fit
 Eq.~(\ref{eq:zptalpha}) in units of the large uncertainty on the
 latter. Assuming the same uncertainty, the better-fit value
 $\alpha_s(M_Z)=0.120$ would instead be compatible with the global best fit
 within uncertainties. 

This result
is at first surprising, as one  might expect that the best fit to the
world data must be along the best-fit line. However, as we we shall show
shortly, it can be understood both at a qualitative, and
also more quantitative level. 

Note that the dataset for the global fit that we are considering actually does
include the $Z$ $p_T$ data of Table~\ref{tab:chi2tabzpt}. Hence, the example
presented here differs somewhat from a standard ``real-life''
situation such as in
Refs.~\cite{Klijnsma:2017eqp}-\cite{Andreev:2017vxu}: there,
PDFs obtained from a fit to a global dataset are 
used for an $\alpha_s$ determination from some new process which was
not among those which were used to determine the PDFs.
In practice, in our case, this makes essentially no difference because the
inclusion of the $Z$
$p_T$ data has almost no  effect on the global fit, due to relatively
small number of data (about a hundred vs. about 4000, see
Table~\ref{tab:chi2tabzpt}), and because the $Z$ $p_t$ data are quite
consistent with other data which determine the same PDFs (essentially
the large $x$ gluon)~\cite{Nocera:2017zge}. This is demonstrated
explicitly in Fig.~\ref{fig:pdfcompnozpt}, where PDFs in the global
fit with or without $Z$ $p_T$ data are compared, and seen to be
essentially identical. Also, $\chi^2$ values for a global fit in which
the $Z$ $p_T$ data are not included are shown in
Table~\ref{tab:chi2tabzpt}, and are seen to be extremely close to
those for the default global fit which includes this data: even the
$\chi^2$ for the $Z$ $p_T$ data themselves are almost unchanged when
fitting this data. We have checked that all $\chi^2$ values for the
other datasets of Table~\ref{tab:chi2taball} change at or below the permille
level upon exclusion of the $Z$ $p_T$ data.

As we will discuss in
Sect.~\ref{sec:zpt} below,  whether or not the data for process $P$
are included in the global fit or not also makes no difference of principle,
though this is besides the point now, given the negligible impact of
the $Z$ $p_T$ data on the global fit.
The reason why we choose
to use for process $P$
dataset which is part of the global dataset,
is that it enables us to use the very large set of 8400 correlated
replicas produced for Ref.~\cite{Ball:2018iqk} in order to construct
the profiles shown in Fig.~\ref{fig:nnlo_parabolas}, thereby ensuring
high statistical accuracy.

We conclude that we have presented an explicit example that shows how,
using an existing PDF set to determine $\alpha_s$ from a particular
process $P$
by looking for the
minimum of the $\chi^2$ for the process along the best-fit line  of the
global fit, can lead to a substantially distorted result. The reason is
that
there
exists  values of $\alpha_s$ for which (for a suitable PDF
configuration) the $\chi^2$ for process $P$ is lower than the minimum
along the best-fit line, but, surprisingly, the $\chi^2$ of the global
dataset is also lower than the value it has at the minimum along the
best-fit line.

This apparently
puzzling result can be qualitatively understood by noting that the
value of $\alpha_s$ which optimizes the $\chi^2$ of the chosen process
is actually closer to the global minimum for $\alpha_s$ than the value
which corresponds to the minimum along the best-fit line. Due to having given
large weight to some process, the
$\chi^2$ for the global dataset deteriorates somewhat, because it is
now optimized for that process, rather than for the global
dataset. But that deterioration is more than compensated by the fact
that the $\alpha_s$ value is now closer to the global minimum. This
is a consequence of the fact that the PDF space is higher-dimensional
(perhaps even infinite-dimensional) so a small distortion of the PDFs
is sufficient to accommodate the highly weighted process, and
consequently the global $\chi^2$ only increases by a small amount due
to the reweighting. In the next section we  cast this qualitative
argument in a more quantitative form.

\begin{figure}[t]
\begin{center}
  \includegraphics[width=0.49\linewidth]{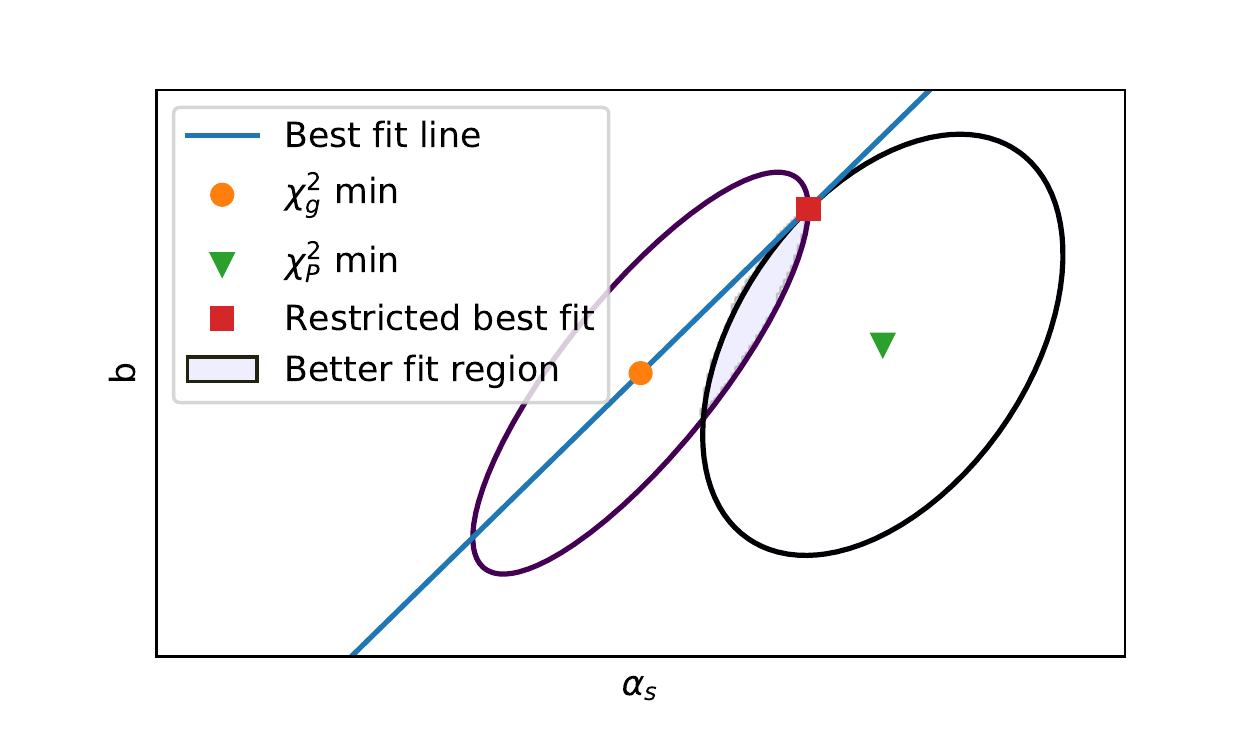}%
  \includegraphics[width=0.49\linewidth]{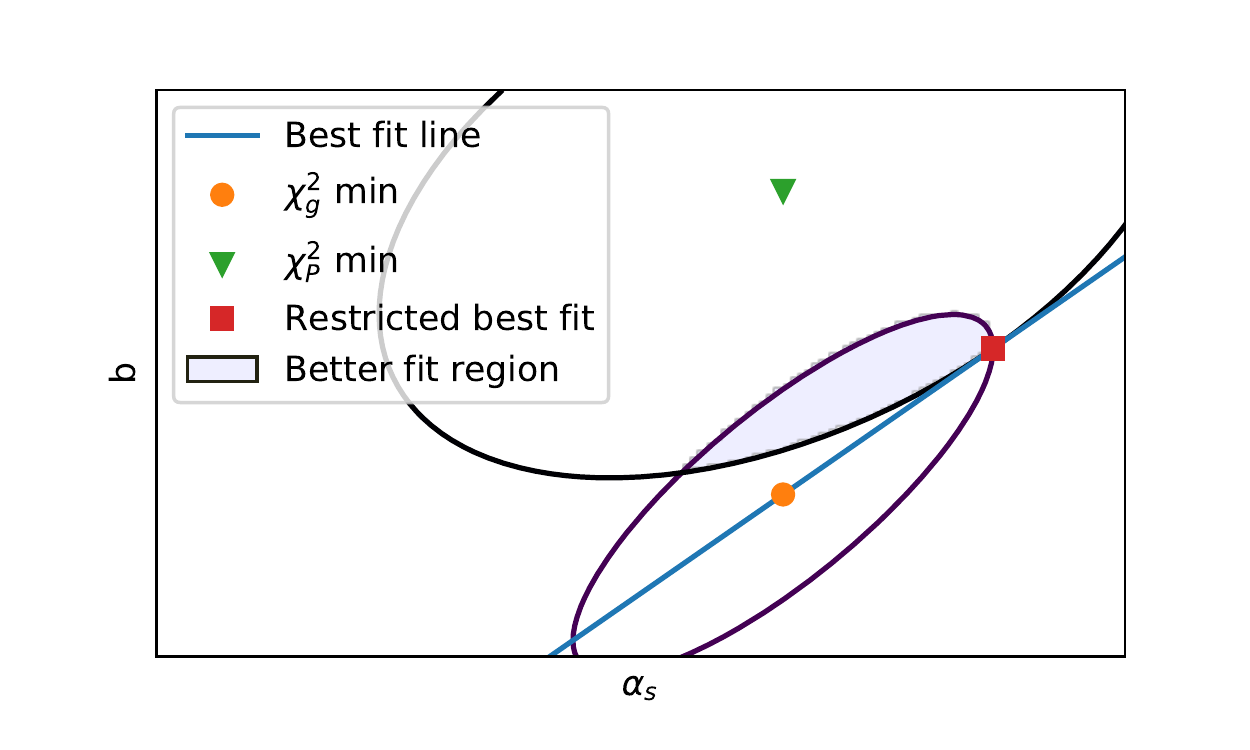}
  \caption{\small Likelihood ($\chi^2$) contours in (PDF, $\alpha_s$)
    space for toy models in which a given process $P$ is sufficient to
    determine PDFs; the parameter $b$ ($y$ axis) schematically
    represents the PDF parameters.
    The minimum of the global $\chi^2_g$ is the orange circle while the
    minimum of $\chi^2_P$ for process $P$ is the green triangle. The
    line is the locus of the best-fit PDF (``best-fit line''): the
    stationary value Eq.~(\ref{bestfitline}) of $b$ for the global $\chi^2$ for fixed
    $\alpha_s$.  The
    red square is the restricted best-fit  $ {\alpha_0^r}^P$: the
    value of $\alpha_s$ corresponding to
    lowest restricted ${\chi^r_P}^2$, i.e. the
    point with lowest  ${\chi_P^2}$ along the best-fit line.
    The ellipses are fixed  $\chi^2_P$ and $\chi^2_g$  contours.
    The shaded area denotes the
    region in which both $\chi^2_g<\chi^2_g( {\alpha_0^r}^P)$ and
    $\chi^2_P<\chi^2_P( {\alpha_0^r}^P)$. The two  plots correspond to
    two possible scenarios (see text).}
 \label{fig:ellipses}
\end{center}
 \end{figure}

\section{The likelihood in (PDF, $\alpha_s$) space}
\label{sec:chisq}

We now discuss some models for the dependence of the likelihood
profiles on $\alpha_s$ and the PDFs which explain the results which we
found in the previous section, and show under which conditions the
situation we encountered can be reproduced. Namely, we explicitly
exhibit likelihood patterns for both a global dataset and a specific
process $P$, such that there exist points in (PDF, $\alpha_s$) space
which have a higher likelihood (lower $\chi^2$) than the restricted
best fit --- the 
point along the global best-fit line in (PDF, $\alpha_s$)
space
which maximizes the likelihood for
process $P$. As in the previous section, we  refer to (minus) the log-likelihood
for the global dataset as $\chi^2_g$, and that for process $P$ as  $\chi^2_P$.

We assume that the global dataset determines
simultaneously the PDFs and $\alpha_s$, so that $\chi^2_g$ 
has a single
minimum value in  (PDF, $\alpha_s$) space, with fixed-$\chi^2_g$ 
ellipses about it. We then consider a particular subset of data,
corresponding to a process $P$: the case of the $Z$ $p_T$ data
discussed in the previous section is an explicit example, but one may
consider both wider datasets (e.g., all LHC data), or smaller datasets
(e.g., one particular measurement of some cross-section performed by
one experiment).

We further distinguish two broad classes of cases. The
first, which is more common, is that process $P$ does not fully
determine the PDFs. This is the case of the $Z$ $p_T$ data of the
previous section, which
constrain the gluon distribution in the  medium-large $x$ range but otherwise
have a limited impact (see in particular Sect.~4.2 of
Ref.~\cite{Ball:2017nwa}). In this case, likelihood contours for
process $P$ in
(PDF, $\alpha_s$) space have flat directions, along which PDFs and $\alpha_s$
change but the value of $\chi^2_P$ does not.
The second is that in which
process $P$ alone is sufficient to provide a determination of the
PDFs, so that  $\chi^2_P$ also has a
minimum in (PDF, $\alpha_s$) space, with fixed-$\chi_P^2$ 
ellipses about it. An explicit example of this would be if process
$P$ was the full set of deep-inelastic scattering data, which do
determine fully the PDFs, albeit with larger uncertainties than a
global dataset~\cite{nnpdflh}. This case is relatively
less common, but we discuss it first because the former case can be
viewed as a spacial case of the latter.

\subsection{Datasets which determine simultaneously $\alpha_s$ and PDFs}\label{sec:sim}
\begin{figure}[t]
\begin{center}
  \includegraphics[width=0.7\linewidth]{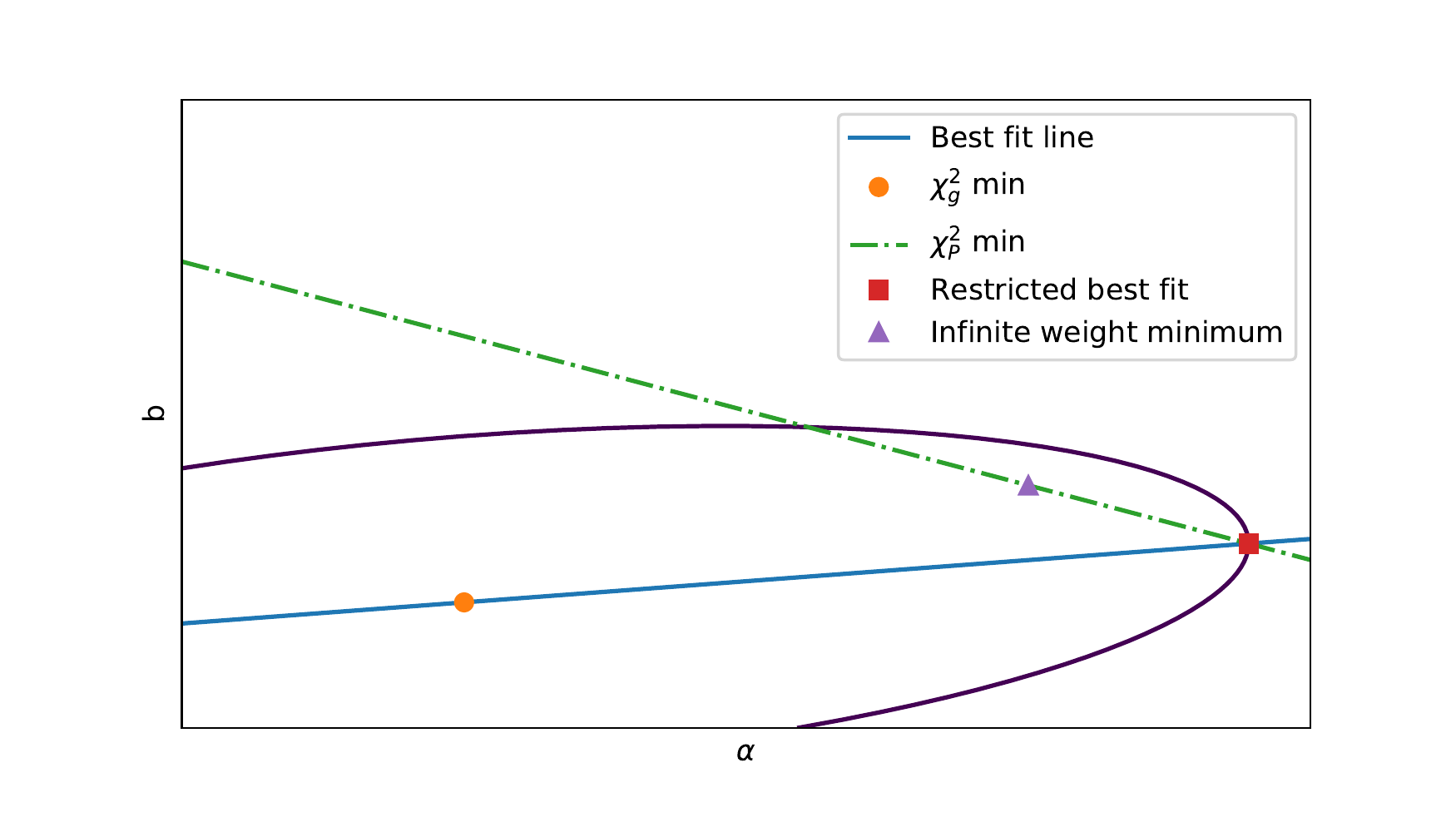}
  \caption{\small Same as Fig.~\ref{fig:ellipses}, but now
    for a toy model in which process P does not fully determine
    the PDFs.
    The minimum of the global $\chi^2_g$ is the orange circle while the
    minimum of the $\chi^2_P$ for process $P$ is the dashed green
    line. The solid blue line is the best-fit line as in Fig.~\ref{fig:ellipses}.
    The
    red square is the ``standard'' value  ${\alpha_0^r}^{P}$:
the
    value of $\alpha_s$ corresponding to
    lowest restricted ${\chi^r_P}^2$, i.e. the
    point with lowest  ${\chi_P^2}$ along the best-fit line.
    The ellipse is a fixed $\chi_g^2$  contour.
    \label{fig:ellipseind}}
\end{center}
\end{figure}

In order to simplify the discussion, we consider a  toy model in which
the whole of  PDF space is represented by a single parameter $b$ so that
(PDF, $\alpha_s$) space is just the two-dimensional ($b$, $\alpha_s$)
plane. In a realistic situation, this can be viewed as a
two-dimensional cross-section of the full space. In the vicinity of
the
minimum, where the $\chi^2$
behaves quadratically,
likelihood contours  are  just ellipses (see Fig.~\ref{fig:ellipses}):
\begin{equation}
\chi_i^2(b,\alpha_s)= \left[\sigma^{i}_1[(\alpha_s-\alpha_0^i)\cos\theta_i
  +(b-b_0^i)\sin\theta_i)]\right]^2+\left[\sigma^{i}_2[-(\alpha_s-\alpha_0^i)\sin\theta_i
  +(b-b_0^i)\cos\theta_i)]\right]^2,\label{chisqcont}
\end{equation}
where $i=g,\>P$ according to whether one is considering the global
dataset, or the dataset for process $P$. In our toy model we neglect the higher-order cubic and
quartic terms that would arise far from the minimum.
The point $(b_0^g,\alpha_0^g)$  (denoted by an orange
circle in Fig.~\ref{fig:ellipses}) corresponds to  the maximum likelihood for
the global dataset, and the point $(b_0^P,\alpha_0^P)$ for
process $P$.

The best-fit line defined in Section~\ref{sec:zpt} is the locus of
points such that
\begin{equation}
  \frac{\partial \chi^2_g}{\partial b}(b,\alpha_s)=0,
  \label{bestfitline}
  \end{equation}
shown
in Fig.~\ref{fig:ellipses} as a (blue solid) line. The
condition Eq.~(\ref{bestfitline}) means that at each
point along this line the tangent to the fixed-$\chi^2_g$ contour is
vertical. Hence, the line is not a principal axis of the ellipse,
unless the principal axes are along the $b$ and $\alpha_s$ directions.
The restricted best-fit point is shown as a red square. This point, $(b^r,
{\alpha_0^r}^P)$, minimizes the restricted  ${\chi^r_P}^2$ along the best-fit
line, so it is tangent to  a fixed $\chi^2_P$ contour.
This is the value of $\alpha_s$ from process
$P$ that
would be determined using the ``standard'' procedure. The value of
$\chi^2$ for process $P$ at this point is the
value discussed in Section.~\ref{sec:zpt}:
${\chi^r_P}^2( {\alpha_0^r}^P)\equiv \chi^2_P(b^r, {\alpha_0^r}^P)$.

The fixed $\chi^2_g$ and $\chi^2_P$ contours
through the restricted best-fit point
are also shown in figure.  It is clear than, whenever they
intersect, the whole area bounded by them (shown as shaded in the
figure) has both $\chi^2_g<\chi_g^2(b^{r}, {\alpha_0^r}^P)$ and
$\chi^2_P<\chi^2_P(b^{r},{\alpha_0^r}^P)$. 
Any point in this
region provides a better fit to both the global dataset and to process
$P$. Whereas it is debatable which $\alpha_s$ value in this region (if any)
should be considered as the best-fit value of $\alpha_s$,  it seems
very difficult to 
argue that the restricted best-fit ${\alpha_0^r}^P$  is
the $\alpha_s$ value preferred by
process $P$, given that it gives a worse fit to the both process $P$,
and the
global dataset than any point in the highlighted region.

The two toy examples shown in Fig.~\ref{fig:ellipses} demonstrate
different cases in which this may happen. Clearly, for some choices of
parameters the
value of the restricted best-fit ${\alpha_0^r}^ P$  might considerably differ from either
of the values $\alpha_0^P$ or  $\alpha_0^g$  that respectively
minimize $\chi^2_P$ or $\chi^2_g$. In fact,
one can exhibit situations, such as shown in the right plot of
Fig.~\ref{fig:ellipses},  in which
$\alpha_0^P\approx  \alpha_0^g$, yet the restricted best-fit  ${\alpha_0^r}^P$ is quite
different. So not only does the restricted best fit provide a worse
fit, but it cannot even be viewed as some kind of average or
interpolation between the global value $\alpha_0^g$ and the process
$P$ value $\alpha_0^P$. This demonstrates that
taking ${\alpha_0^r}^P$ as the value of $\alpha_s$ determined by
process $P$ leads to an incorrect result.

\subsection{Datasets which do not fully determine the PDFs}
\label{sec:insuf}

We now turn to the case in which process $P$ does not fully determine
the PDFs, so that there are flat directions for $\chi^2_P$ in (PDF,
$\alpha_s$) space. This means that, whereas the
likelihood profile for the global dataset still has the form of
Eq.~(\ref{chisqcont}), for process $P$ there exists a hypersurface
in (PDF, $\alpha_s$)
space (i.e. in our toy model a curve in the $(b,\alpha_s)$ plane)
along which
$\chi^2_P$ is at a minimum. This can be viewed as a limiting
case of Eq.~(\ref{chisqcont}), when the fixed $\chi_P^2$ ellipses  become infinitely thin, i.e., when
either of $\sigma^{P}_i$ goes to zero. Of course, just like far enough
from the minimum the fixed-$\chi^2$ profile will no longer be
ellipsoids, the flat direction will only be locally  straight. 
This situation is depicted in
Fig.~\ref{fig:ellipseind}, where the minimum curve for
$\chi^2_P$ is
shown as a (dashed, green) straight line. In this case, in the generic
situation in which this minimum curve and the best-fit line
Eq.~(\ref{bestfitline}) intersect, the intersection point is the
restricted best fit
$(b^{r},{\alpha_0^r}^P)$,
which would provide the ``standard''
$\alpha_s$ determination. 

However, it is clear that if one now considers the fixed $\chi^2_g$
contour through this point (shown as the ellipse in
Fig.~\ref{fig:ellipseind})
in a generic case, i.e. unless the
minimum curve (the dashed green curve of Fig.~\ref{fig:ellipseind}) is
tangent to this ellipse, the contour intercepts a segment of the
minimum curve, and any point along this segment  provides a better
fit to the global dataset than the restricted best-fit  $(b^r,{\alpha_0^r}^P)$.
The minimum of the global $\chi^2_g$ along this segment is shown as a purple triangle in
Fig.~\ref{fig:ellipseind}. Clearly,
this is the point that is selected by minimizing the weighted
\begin{equation}
  \chi^2_w=\chi^2_g+ w \chi^2_P
  \label{weighted}
\end{equation}
in the limit of very large $w$. Indeed, in the limit in which $w$ is
very large so $w\chi^2_P\gg \chi^2_g$ the minimum of  $\chi^2_w$ is
along the line of degenerate minima of 
$\chi^2_P$, but for any finite $w$ the absolute minimum of $\chi^2_w$
is at the point at which $\chi^2_g$ is also minimal. 

Arguably, the value of $\alpha_s$ at this large-weight minimum  can be viewed as the  best-fit value
$\alpha_0^{P}$ of
$\alpha_s$ as determined from  process $P$,
subject to the constraint of also fitting the global dataset. 
Be that as it may, the  best-fit value of
$\alpha_s$ as determined from  process $P$ is surely not the
restricted best-fit ${\alpha_0^r}^{P}$, which leads to a worse fit to
the global dataset than any value of
$\alpha_s$ along the intercept segment.

This is then representative of the case that we discussed in the
Section~\ref{sec:zpt}. On the one hand, the value
${\alpha_0^r}^{P}$ does not generically provide the best simultaneous
fit of process $P$ and the global dataset. Also, the value
that minimizes the weighted $\chi^2$ for large $w$ --- which provides
a better fit to the global dataset while giving a fit of the same
quality to process $P$ --- is generally closer to the global
best-fit $\alpha_0^g$, as it is clear from
Fig.~\ref{fig:ellipseind}.
Note that in this
simple example, in which PDF space is one-dimensional,
the large-$w$ minimum leads to the
same fit quality for process $P$ as the restricted minimum. In a
realistic situation  both
flat and non-flat directions will be present, and the
weighting will also change the position of the minimum along the
non-flat direction, thereby leading to a lower $\chi^2$ for process
$P$ than the restricted minimum, as we observed in Section~\ref{sec:zpt}.

We conclude that the situation we encountered in Section~\ref{sec:zpt}
is generic. Whenever process $P$ does not fully determine the PDFs,
$\chi^2_P$ in (PDF, $\alpha_s$) space has a
subspace of degenerate minima.
The value of $\alpha_s$ obtained by minimizing the restricted
${\chi^r_P}^2$  then leads to an incorrect result,
generally further away from the global best-fit $\alpha_0^g$ than the
value that would be obtained by looking for the minimum of the global
$\chi^2_g$ in this subspace of degenerate minima of $\chi^2_P$.

It is important to note that this effect can be quite large, as it was
the case in the explicit example of the previous section. In general,
the size of the deviation of the  infinite weight minimum  from the
restricted minimum will depend on the numerical values of the
parameters that characterize $\chi^2_g$ and $\chi^2_P$
Eq.~(\ref{chisqcont}). Note however that whenever the restricted best fit
differs considerably from the global best fit in units of the standard
deviation of the global best fit, then the $\chi^2_g$ parabola will vary rapidly
in the vicinity of the restricted best fit, and thus the infinite
weight minimum will generically have a rather different value. This
is the case of the example of Section~\ref{sec:zpt}, in which the
restricted minimum Eq.~(\ref{eq:zptalpha}) is eleven standard
deviations away from the global minimum Eq.~(\ref{eq:globalpha}). It
is interesting to observe  that in the recent determination of
$\alpha_s$~\cite{Ball:2017nwa} many of the restricted minima from
individual datasets indeed differ considerably from the global
minimum. 

As a final observation, we note that the argument presented
here, and thus its conclusion, are unaffected
regardless of whether process $P$ is or is not included in the global
dataset. This has the interesting implication that in a global
simultaneous determination of $\alpha_s$ and the PDFs, such as
performed in Ref.~\cite{Ball:2018iqk}, the minimum of
$\chi^2$ from each dataset entering the global determination cannot be
interpreted as the $\alpha_s$ value corresponding to that
dataset. Hence, there is no reason to expect that the global best-fit $\alpha_s$ is the
mean of the restricted best-fit values determined from each subset of
the data entering the global fit.

\section{The value of $\alpha_s$ from a single process} 
\label{sec:concl}

The main conclusion of this paper is that it is generally not possible
to reliably determine $\alpha_s$ from a given physical process which
depends on parton distributions while relying on  a pre-existing PDF
set. The reason can be simply stated: the existing PDF sets only
sample a line in PDF space as $\alpha_s$ is varied, hence, when using
them, one is determining a constrained likelihood of the physical process
under investigation along this line. This biases the results of the
determination, in 
that the true maximum likelihood $\alpha_s$ 
generally corresponds to a PDF configuration which is not along this
line. The bias is especially severe since PDF space is
high-dimensional.
We have proven our point by showing that there exist PDFs
which provide a better fit both to the given process,
and the global dataset, and correspond to a different $\alpha_s$ value.
This has been shown both in an explicit example, and in toy models.
Interestingly, when the  physical process
under investigation does not fully determine the PDFs, we have shown that
this bias will generically pull the value of $\alpha_s$ away from the best
fit, in comparison to values of $\alpha_s$ which provide a better fit to both
the given process and the global
dataset. Hence, determining $\alpha_s$ from individual processes in
this way, artificially inflates the dispersion of the $\alpha_s$
values which are found.

It is important to stress that the problem that we are pointing
out cannot be viewed as an extra source of PDF uncertainty in a
determination which uses a pre-existing PDF set, but rather, it
exposes a conceptual flaw. Indeed, the value of $\alpha_s$ found by
not fitting the PDF simultaneously does not correspond to a
maximum likelihood point in (PDF, $\alpha_s$) space, and as such it
can differ from the true maximum likelihood point
by an amount which is potentially large (as we have shown in explicit
examples),
and impossible to quantify
without knowledge of the PDF dependence of the results.

One may then ask: what is the value of $\alpha_s$ determined by
process $P$? Does it exist at all?
Clearly, in
the case in which the dataset for process $P$ is wide enough that it can be
used to simultaneously determine both  $\alpha_s$ and the PDFs, it is this
value of $\alpha_s$ which must be interpreted as the value preferred
by process $P$. In this case, the main import of our
analysis is to  show that
minimizing along the line of global best-fit PDFs may lead to a  value of
$\alpha_s$ which not only provides a poor fit to both process $P$ and
the global dataset, but cannot even be viewed as some kind of average
of the value $\alpha_0^{P}$ from process $P$ and the global value
$\alpha_0^{g}$; rather, it will randomly differ from them in a way which
depends on the $\chi^2$ profiles in (PDF, $\alpha_s$) space (see the
right plot in Figure~\ref{fig:ellipses}).

On the other hand, it is very common that the process $P$ is insufficient to
simultaneously determine $\alpha_s$ and the PDFs, and hence for $\chi^2_P$ to
have a set of degenerate minima in (PDF, $\alpha_s$) space. In this case it
is debatable whether it makes sense to speak of a value of $\alpha_s$
determined by process P. One may take the purist attitude that such value
does not exist, or, alternatively consider defining the best fit value
of $\alpha_s$ as the result of the weighting procedure discussed in in
Section~\ref{sec:insuf}, i.e., as the best fit to the global dataset
within the set of degenerate minima of the $\chi^2_P$.
 In such case, the
uncertainty on this $\alpha_s$ value is  determined by conventional
one-$\sigma$ contours of the global $\chi^2$ in the degenerate subspace
(i.e., in the example of Fig.~\ref{fig:ellipseind}, along the dashed green
line). 

The important
observation in this case  is that the value found minimizing along the best-fit
line will generally be further away from  the global best fit, while
providing a worse fit to both process $P$ and the global dataset.  So
in particular if one wishes to assess the spread of values of
$\alpha_s$ which are individually favored by each of the individual
processes which enter in a global simultaneous determination of PDFs
and $\alpha_s$ (such as that of Ref.~\cite{Ball:2018iqk}) a 
realistic estimate is found by weighting each of the individual
datasets in turn, while the spread of the restricted minima will
suggest an artificially inflated dispersion of values.

The upshot of this whole discussion is that we do not envisage a shortcut:
a determination of $\alpha_s$ from a single process always 
requires a simultaneous determination of  PDFs. In the simplest
case, of a process (such as deep-inelastic scattering) which is
sufficient to determine the PDFs, one must perform a simultaneous fit
of the PDFs and $\alpha_s$ to the dataset for that process. In the
more common case of a process which does not fully determine the PDFs
one may determine a value of $\alpha_s$ for this process (if deemed
interesting) through the weighting method discussed above, but this of
course requires performing anyway a global PDF fit: so it is no easier
than simply including process $P$ in the dataset
and repeating the global simultaneous determination of the PDFs and
$\alpha_s$.

In this latter case, of performing a global fit of PDFs and
$\alpha_s$, it might at least in principle be
possible to include the new
dataset, without refitting, by Bayesian
reweighting\cite{Giele:1998gw,Ball:2010gb}.
Indeed, there is no difficulty of principle in reweighhting 
correlated replicas: each replica will then correspond not only to a
different set of PDFs, but also to a different $\alpha_s$ value (that
given by Eq.~\ref{eq:alphakdef}). The reweighted replica ensemble then
also gives a posterior distribution of $\alpha_s$ values. Whether and
how the procedure would work 
 when the new dataset is given a large weight is however not
 immediately clear. Also, whether
this is feasible in practice of course remains to be seen:
specifically, it might well be that in  concrete cases an
unrealistically large number of  replicas in the prior set is  necessary in
order to get a reliable answer after reweighting.

Our results have two wider sets of implications. On the one hand, they
provide a strong indication that looking at the $\chi^2$ profile for any given
process in the subspace of global fits as one parameter is varied can
be very misleading. This is true not only for $\alpha_s$ but for any
parameter entering the global fit, including the parameters which
govern the shape of the PDF themselves. Specifically, the dispersion
of best-fit minima for individual processes as a feature of the PDF is
varied
-- such as, say, the rate at which the gluon grows at small $x$ --
does not appear to be a good proxy of the actual dispersion of the
results favored by each processes. This may have some relevance in the
benchmarking of parton distributions (see e.g.
Refs.~\cite{Ball:2012wy,Andersen:2014efa}).
   
On the other hand, they suggest caution in the determination of any
standard model parameter from hadronic processes.
Indeed, while the case of the determination of $\alpha_s$ is particularly
relevant because of the very strong correlation of $\alpha_s$ and the
PDFs, similar considerations apply to the simultaneous determination
of any physical parameter in PDF-dependent processes, such as the
determination of the top quark mass mass~\cite{Alekhin:2016jjz},
or of electroweak parameters, such as the $W$ mass~\cite{Bagnaschi:2019mzi}. In
the latter case, the correlation of PDFs and the parameter is in
principle weaker than in the case of the strong coupling,
but the very high accuracy which is sought suggest that currently available
results, specifically in $W$ mass determination, should be
reconsidered with care.

\section*{Acknowledgments}
We thank the members of the NNPDF collaboration for several
discussions, in particular 
Richard Ball, whom we also thank for a critical reading of the
manuscript, and Rabah Abdul Khalek for comments.
ZK thanks G.~Salam for interesting discussion and stimulating questions.\\
ZK is supported by the European Research Council Consolidator Grant
``NNLOforLHC2'' (n.683211).
SF is supported by the European Research Council under
the European Union's Horizon 2020 research and innovation Programme
(grant agreement n.740006).

\bibliographystyle{UTPstyle}
\bibliography{noalpha}

\end{document}